\documentclass[%
 reprint,
 amsmath,amssymb,
 aps,
]{revtex4-2}

\usepackage{graphicx}
\usepackage{physics}
\usepackage{color}
\usepackage{amsmath}
\usepackage{amsfonts}
\usepackage{xspace}
\usepackage{array}
\usepackage{makecell}
\usepackage{graphicx}
\usepackage{dcolumn}
\usepackage{bm}

\usepackage{ulem}
\newcommand{\red}[1]{{\color{red}#1}}

\def\be{\begin{equation}}
\def\ee{\end{equation}}
\def\bea{\begin{eqnarray}}
\def\eea{\end{eqnarray}}

\begin{document}

\preprint{APS/123-QED}

\title{Quantum Many-body Scar Models in One Dimensional Spin Chains} 

\author{Jia-Wei Wang}
\author{Xiang-Fa Zhou}
\email{xfzhou@ustc.edu.cn}
\author{Guang-Can Guo}
\author{Zheng-Wei Zhou}
\email{zwzhou@ustc.edu.cn}
\affiliation{CAS Key Lab of Quantum Information, University of Science and Technology of China, Hefei, 230026, China}
\affiliation{Synergetic Innovation Center of Quantum Information and Quantum Physics, University of Science and Technology of China, Hefei, 230026, China}
\affiliation{ Hefei National Laboratory, University of Science and Technology of China, Hefei 230088, China}
\date{\today}	

\begin{abstract}
The phenomenon of quantum many-body scars has received widespread attention both in theoretical and experimental physics in recent years due to its unique physical properties. In this paper, based on the $su(2)$ algebraic relations, we propose a general method for constructing scar models by combining simple modules.
This allows us to investigate many-body scar phenomena in high-spin systems. We numerically verify the thermalization and non-integrability of this model and demonstrate the dynamical properties of the scar states. We also provide a theoretical analysis of the properties of these scar states. For spin-$1$ case, we find that our 1D chain model reduces to the famous PXP model[C. J. Turner et al. Phys. Rev. B 98, 155134(2018)] under special parameter condition.
In addition, due to the continuous tunability of the parameters, our model also enables us to investigate the transitions of QMBS from non-integrable to integrable system.
\end{abstract}

\maketitle


\section{Introduction}
In recent years there has been a main interest in quantum many-body scar phenomena(QMBS), which was first discovered experimentally in 1D Rydberg atom chain model in 2017~\cite{quantumscarexp_2017}.
Under appointed initial states, such systems with QMBS will lead to violation against Eigenstates Thermalization Hypothesis(ETH)~\cite{eth1_1991,eth2_1994,eth3_2011,eth4_2016} and show periodical revival of the initial state as Loschmidt echo.
Such 1D Rydberg atom chain with blockade interaction, that prohibits adjacent Rydberg excited states, is named PXP model~\cite{cjturner1,cjturner2}.
To explore its eigenstate structure, a large amount of research has been carried out such as forward scattering approximation~\cite{cjturner1}, hiring spectrum generating algebra~\cite{pxpsga,scarforall}, composite spins~\cite{zhaihui}, Hilbert space fraction~\cite{fractionpxp,Moudgalyafragment,fragmentation2} etc. Besides, other models causing QMBS, including Onsager’s scars~\cite{onsager}, Fermion-Hubbard model~\cite{fermihubbard}, Bose-Hubbard model~\cite{bosehubbard}, integer spin AKLT models~\cite{AKLT1,AKLT2}, spin-$1$ XY model~\cite{XY}, 1D spin-$1$ Kitaev model~\cite{spin1kitaevmodel}, coupled top model~\cite{topmodelqmbs}, truncated Schwinger model~\cite{Schwingermodel}, etc., have also been studied.

Although the phenomenon of quantum scars has been extensively studied in 1D low-spin systems (such as spin-$1/2$ and spin-$1$ chains)~\cite{cjturner1,projection}, there has been relatively less researches on high-spin systems~\cite{QMBSsyzh,highspinPXP}. Comparatively, high-spin systems have more internal degrees of freedom, making it more challenging to construct scar states using constrained interactions.
It is thus interesting to find which type of constraints or blockade interactions possess the possibility of causing thermalized phenomena in high-spin chains, but still support the existence of quantum scar states.
It is also remains unclear whethere these Hamiltonians can be obtained by appropriately combining simple solvable modules.
On the other hand, exploring the evolution of quantum scar states as a system transitions from integrable to non-integrable (with QMBS), is also an important means of understanding the phenomena of thermalization.
However, currently, there are relatively few relevant theoretical models that support this condition~\cite{projection,scarforall,breakexactscar}.
Therefore, finding and constructing effective interactions that meet this condition is another important issue that we are concerned with.

In this paper, we consider constructing models that support QMBS in 1D systems using simple modules that satisfy the $su(2)$ algebraic structure.
This is achieved by averaging two effective collective spin-$j$ operators, which are mutually connected through a mirror reflection defined by nonlocal unitary transformations.
Unlike the former works in which QMBS phenomena emerged in isolation with fixed Hamiltonian, we have been able to come up with scar models and 'scar-like' models constructed from basic building blocks satisfying the  $su(2)$ relation. 
Our strategy can be generalized to construct 1D models in high-spin systems that host fragmentation and QMBS, and covers previous QMBS models in ~\cite{cjturner1,highspinPXP} as special cases.
Furthermore, by continuously varying parameters, our models can be transitioned to the non-interacting cases, which reveals the inherent relation between QMBS eigenstructure and the $su(2)$ algebraic structure.

The paper is organized as follows. In section II, initiated with a 1D non-interacting spin chain, we construct the QMBS model by introducing unitary transformations based on $su(2)$ algebra and defining the Hamiltonian as the weighted average of two collective spin operators. Then the models brought up are shown to host QMBS phenomena in section III. Next, in section IV, we study the blockade scenario with a specifically chosen parameter in the Hamiltonian. The close relation of the present model in case of $j=1$ with the well-known PXP model is also analyzed in detail. In section V, a systematic approach to approximating the scar eigenstates is then provided and verified both analytically and numerically.
In section VI, thanks to the tunability of the model, we have also investigated the transformation of the scar states as the system undergoes a transition from non-integrability to integrability. We conclude this work in the final section.

\section{Quantum Scar Models Based on $su(2)$ Algebra in One-dimensional Spin Chains}


We start by constructing a series of models supporting QMBS in one-dimensional chains with large spin.
For an $N$-site spin-$j$ system, we consider the collective spin operator defined by $J^{\alpha}=\sum_{l=1}^{N}S_l^{a}$.
Here $\alpha=\{x,y,z\}$, $N$ indicates the total number of the spins, and $S_l^a$ represents the spin-$j$ operator acting on $l$-th lattice site. These collective operators form the algebraic structure of $su(2)$ and satisfy the corresponding commutation relation  $[ J^a ,J^b ]=i \epsilon_{abc} J^c$, where $\epsilon_{abc}$ represents the component of the antisymmetric Levi-Civita tensor with $\epsilon_{xyz}=1$.

To construct the desired scar models, we then introduce effective blockade characteristic interactions between the nearest neighbor spins. This is achieved by introducing an unitary transformation $U(\theta)=e^{i\theta \hat{C}}$, where the Hermitian operator $\hat{C}$ is defined as $\hat{C}=\sum_{l=1}^{N} \Pi_l^{(j)}\otimes \Pi_{l+1}^{(-j)}$,  which acts on the $l$-th and $(l+1)$-th spin and counts the numbers of the patterns $\left| j,-j\right\rangle_{l,l+1}$ in the chain. Here $|m\rangle_l$ represents the local spin eigenstates of $l$-th site satisfying $S^z_l|m\rangle_l=m|m\rangle_l $, $\Pi_l^{(m)}=| m\rangle \langle m|_l$ represents the projection onto the local spin state $|m\rangle_l$, and $\left| j,-j\right\rangle_{l,l+1}=|j\rangle_l\otimes|-j\rangle_{l+1}$ is the tensor product of neighboring local states $\left| j \right\rangle_l$ and $\left| -j \right\rangle_{l+1}$. Throughout the entire work, we employ periodic boundary condition(PBC) for convenience.
The explicit form of $U(\theta)$ can be rewritten as
\begin{equation}\label{lll}
U(\theta)=\prod\limits_{l=1}^{N}\left[ \mathbb{I}-(1-e^{i\theta})\Pi_l^{(j)}\otimes \Pi_{l+1}^{(-j)} \right],
\end{equation}
with $\theta$ satisfying $U(\theta+2\pi)=U(\theta)$. Here $\mathbb{I}$ is the identity matrix of the entire spin chain.
Especially, when $\theta=\pi$, we have the reflection transformation $U_{\pi}\equiv U(\pi)=\prod_{l=1}^{N}\left[ \mathbb{I}-2\Pi_l^{(j)}\otimes \Pi_{l+1}^{(-j)} \right]$ and $U_{\pi} U_{\pi}=\mathbb{I}$, which represents cascade flip operation along the hyper-planes defined by $\Pi_l^{(j)}\otimes \Pi_{l+1}^{(-j)}$.  Using this transformation, the Hamiltonian of the model can then be written as $H(\theta,a)=U(\theta)H(a)U(-\theta)$, with $\theta \in [0,2\pi)$ and $H(a)$ being the weighted average of two trivial terms
\begin{equation}\label{eqhamiltonian}
\begin{aligned}
 H(a)=\frac{1+a}{2}J^x(0) +\frac{1-a}{2}J^x(\pi),
\end{aligned}
\end{equation}
where $a$ stands for the averaging weight satisfying $a\in[0,1]$, and $J^{\alpha}(\theta)=U(\theta)J^{\alpha}U(-\theta)$ denotes the deformed collective $su(2)$ generators defined by $J^{\alpha}$ and the unitary transformation $U(\theta)$.

\begin{figure}
	\centering
	\includegraphics[width=7cm]{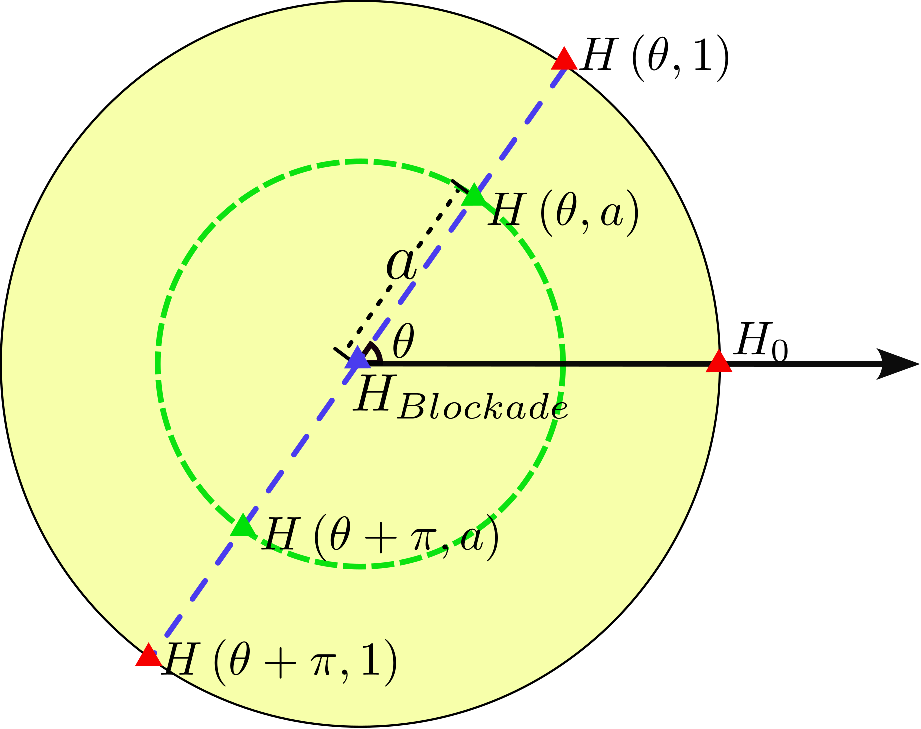}
	\caption{Illustration of the Hamiltonians introduced here, which are distributed within a circle of radius $1$ on the parametrical plain. There are several different categories of Hamiltonians, including the original one $H_0=H(0,1)$, $H(\theta,1)$ obtained by unitary transformation (marked with red triangles), the blockade one $H_{Blockade}=H(\theta,0)$ (marked with a blue triangle) and a sample of general case $H(\theta,a)$ (marked with green triangles).}
	\label{fig: Demonstration of the parametrical distribution of hamiltonian}
\end{figure}

We note that the above construction represents a typical method to obtain a spin model supporting QMBS starting from simple collective operators.
To illustrate the symmetry of the models, in Fig.~\ref{fig: Demonstration of the parametrical distribution of hamiltonian}, we represent all these Hamiltonians $H(\theta,a)$ on the parametric  $\theta-a$ plane with $\theta$ and $a$ being the polar angle and radius respectively.
These Hamiltonians form a unit circle on the plane.
The Hamiltonian can also be rewritten as 
\begin{eqnarray}\label{equnfoldedH}
&&H(\theta,a)=\sum_{l=1}^{N}S_l^x+ \nonumber \\
&& \hspace{.95cm} \sqrt{\frac{j}{2}}
\left \{ [ (c-1)|j\rangle \langle j-1|_l +h.c. ] \otimes \Pi_{l+1}^{(-j)} \right.\\
&& \hspace{1.7cm} + \left. \Pi_l^{(j)}\otimes  [ (c-1)|-j\rangle \langle -j+1|_{l+1} +h.c.] \right \} ,  \nonumber
\end{eqnarray}
where $c=a e^{i\theta}$ is the complex number on the parametric plane. On this plane the Hamiltonians with the same radius differ from each other only by a unitary transformation $U(\theta_1-\theta_2)$. The first term on the left in Eq.(\ref{equnfoldedH}) is the non-interacting spin Hamiltonian, and the last two terms represent the interaction between adjacent lattice sites whose strength is determined by the parameter $c$.
Specifically, the spin flip $j\leftrightarrow j-1$ ( or $-j+1\leftrightarrow -j$) occurs on lattice site $l$ only when its neighboring site $l+1$ (or $l-1$) is set to be $|-j\rangle_{l+1}$ (or $|j\rangle_{l-1}$ ). 
This indicates that nearest-neighbor interactions only occur on specific minority states.
It is thus expected that as the increase of the spin size $j$, the subspace affected by the interaction will become smaller relative to the entire system. The evolution of the system will be predominately governed by the first term without interactions.
Since $H(\theta,a)$ is equivalent to $H(a)$ up to a unitary transformation, the eigen-structure of $H(\theta,a)$ is completely the same as $H(a)$. In the following discussions, we will mainly focus on $H(a)$ for simplicity.

\begin{figure}
	\centering
	\includegraphics[width=8cm]{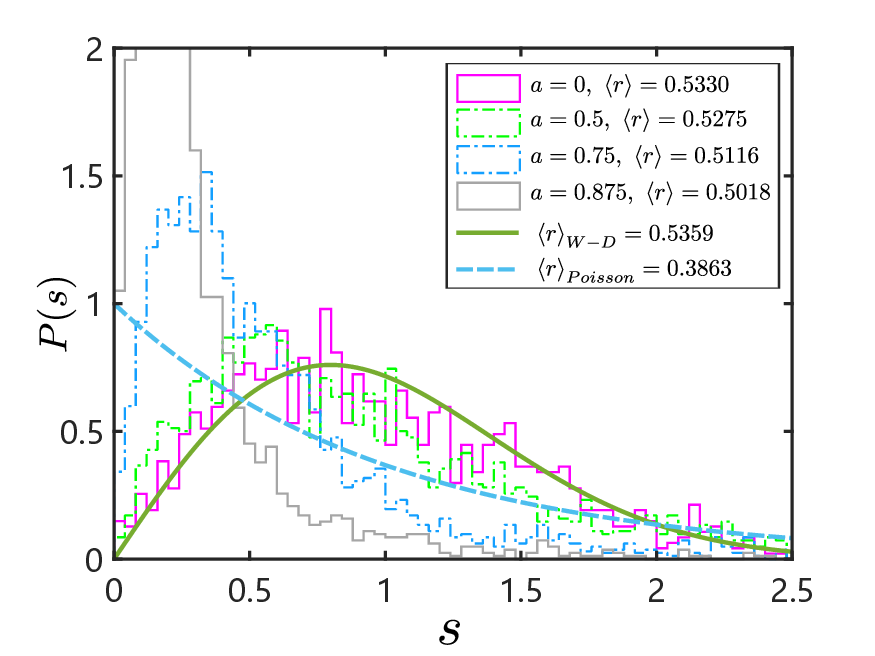}
	\caption{The probability distribution of energy spacings for a spin chain with spin-$\frac{3}{2}$ and $N=8$. The parameter $\theta$ is kept unchanged while $a$ is tuned as $a=1-\frac{1}{2^n}$. The corresponding averaging of $r_i$ is demonstrated in the inner box on the right upper corner.}
	\label{fig: calsforrmt}
\end{figure}

\section{Numerical verification of quantum many body scar phenomenon}

The continuous distribution of Hamiltonians on the plane allows us to discuss the scar behaviors under different parameters. When $a=1$, since only one term is left in $H(a)$, the model is exactly solvable. However, since the two terms in $H(0)$ are equally weighted for $a=0$, $H(0)$ become non-integrable, and show the properties of QMBS. In the latter case, the system possesses an additional symmetry, and the total Hilbert space is decomposed into disconnected subspaces under a similar mechanism of 'weak fragmentation' as discussed in \cite{Moudgalyafragment}.

To illustrate the non-integrability of the system, we calculate the energy spacings of the models' Hamiltonian in comparison with the standard Wigner-Dyson and Poisson distributions. Here the energy spacing is defined as $s_l=S_l/\overline{S_l}$ with $S_l=E_l-E_{l+1}$, where $E_l$'s are sorted energy levels and $\overline{S_l}$ represents mean energy spacing in the vicinity of $E_l$. It is demonstrated in Fig.~\ref{fig: calsforrmt} that when $a=0$, $P(s)$ fits the typical Wigner-Dyson distribution which signifies the non-integrability of the model. However, when $a$ approaches $a=1$, the clear Wigner-Dyson distribution at $a=0$ gradually merges into a non-chaotic case where the distribution of energy spacings peaks near zero.
In the latter case, the system becomes integral. Considering the fact that the average \red{$\left\langle r\right\rangle$} of $r_l=\min(s_l,s_{l+1})/\max(s_l,s_{l+1})$ does not experience a sharp drop, the system can no longer be treated with random matrix theory anymore~\cite{rmt,rmtbook}.

\begin{figure}
	\centering
	\includegraphics[width=\linewidth]{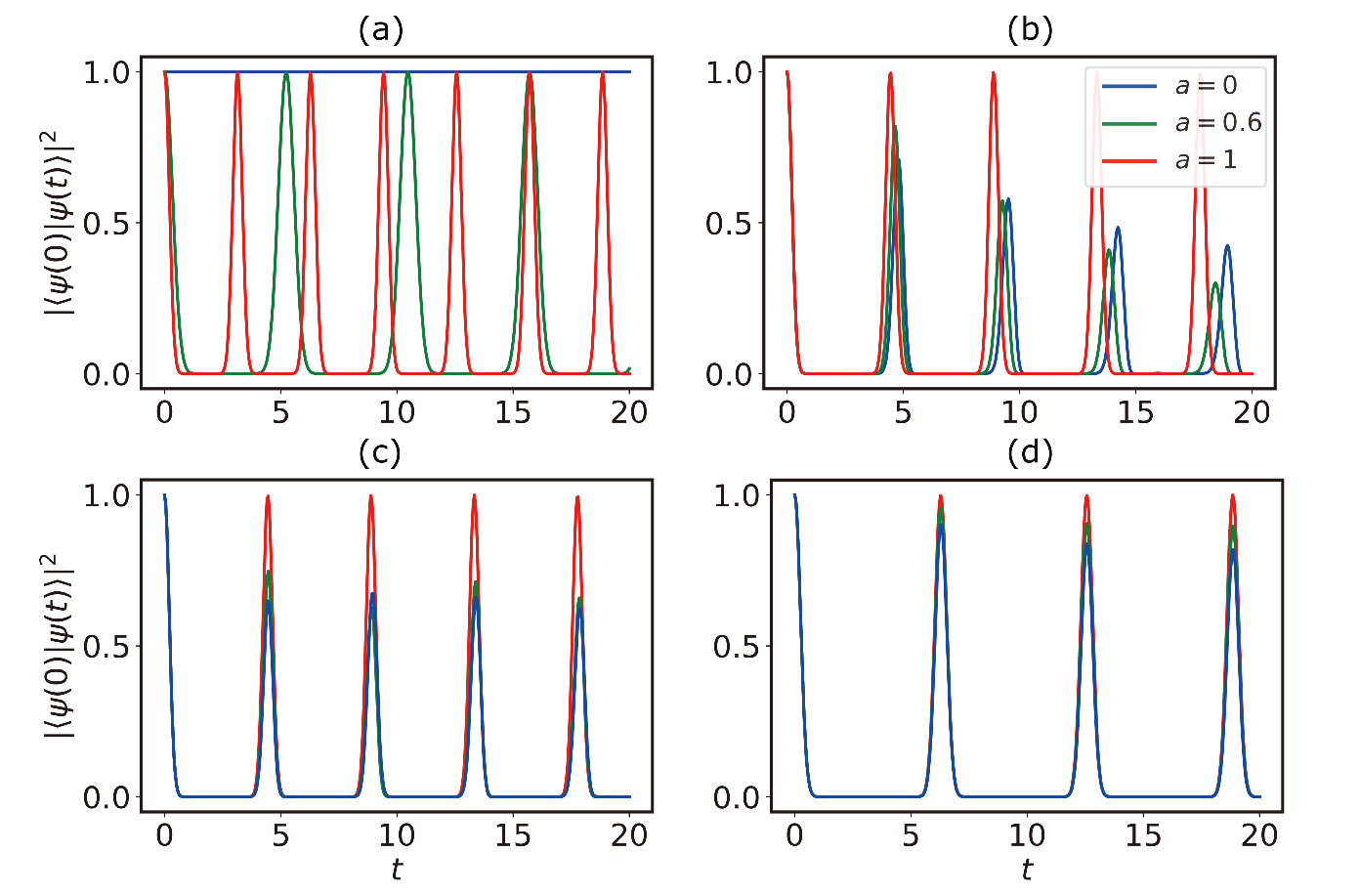}
	\caption{Periodical revival of initial states for the scar models shown in Eq.(\ref{eqhamiltonian}) with lattice size $N=12$ in (a) and (b), where the spin size are set to be $j=\frac{1}{2}$ and $1$ respectively. For larger spin size  $j=\frac{3}{2}$ and $2$ in (c) and (d), the lattice size is set to be $N=8$. Here blue, green, and red lines separately correspond to the parameter $a=0,0.6,1$ in the Hamiltonian with fixed $\theta = 0$. It can be seen that the periodical revival with damping amplitude emerges in spin-$1,\frac{3}{2},2$ models, which is a rather iconic feature for many-body systems supporting QMBS. However, in the special case with  $j=\frac{1}{2}$,  the revivals are always precise.}
	\label{fig: periodicalrevival}
\end{figure}

As one of the critical characteristics verifying the existence of QMBS, we calculate the evolution of the fidelity $|\langle \psi(t)|\psi(0)\rangle |^2$ with the chosen initial state $\left| \psi(0)\right\rangle=\bigotimes_{l=1}^N| j\rangle_l$.
The numerical results show that $|\langle \psi(t)|\psi(0)\rangle |^2$ exhibits the usual collapse-recovery phenomenon as plotted in Fig.~\ref{fig: periodicalrevival} when the spin size $j\geq 1$ is taken.
However, this dynamical behavior deviates from the perfect periodic motion, as the amplitude of recovery gradually decreases in the evolution process, and the peak position also does not exhibit exact periodicity.
It is noted that all the models within the entire circle shown in Fig.~\ref{fig: Demonstration of the parametrical distribution of hamiltonian} support quasi-periodical revival, which becomes more and more rigorous when the spin size $j$ becomes larger.
Physically, when $j$ increases, the proportion of the interaction part in the system becomes relatively smaller. Therefore, the dynamics of the system will increasingly resemble a non-interacting spin system, and thus tend to exhibit an ideal periodic behavior.

Here, we emphasize that the scar phenomena appear only when the spin size $j\geq1$ in our model. In the case of $j=1/2$, we observe static dynamics when $a=0$, and perfect periodic revivals when $a\geq 0$, which can not be characterised as QMBS, as shown in section IV and Appendix C. In the case of $j=1,a=0$, the model can be mapped to PXP model, as will be shown in section IV.

\section{QMBS and the blockade Hamiltonian with $a=0$.}
Having demonstrated the scar phenomena by showing the periodic revival of initial states in Fig.~\ref{fig: periodicalrevival}, we restrict ourselves to Hamiltonian $H(0)$ at the origin of the parametric plane with $a=0$.
The explicit form of $H(0)$  can be written as
\begin{equation}\label{a0Hamiltonian}
H(0)=\frac{J^x(0) +J^x(\pi)}{2},
\end{equation}
which remains unchanged under the unitary transformations $U(\theta)$.
This indicates that the Hamiltonian commutes with the counting operator $\hat{C}$ as $[ H(0),\hat{C} ]=0$.
Therefore, the eigenspace fragments into different subspaces labeled by the eigenvalues of the operator $\hat{C}$, which counts the number of the patterns $\left| j,-j \right\rangle_{l,l+1}$.
Each subspace with $C\neq 0$ then further fragments into multiple smaller subspaces with the location of $\left| j,-j \right\rangle_{l,l+1}$ patterns fixed in the spin chain.
This fragmentation appears to be a weak one which only violates the strong ETH, as opposed to the cases caused by dipole moment conservation in the pair hopping model, $t-J_z$ model, etc~\cite{fractionpxp,fragmentation2,Moudgalyafragment}.
The $C=0$ subspace is the biggest fracture in the entire eigenspace with all the patterns $\left| j,-j \right\rangle_{l,l+1}$ forbidden in the subspace.
In this case, the interaction term in Eq.(\ref{equnfoldedH}) corresponds to a nearest neighbor blockade term added to the non-interacting system, which prohibits adjacent spins from occupying the states $\left| j, - j \right\rangle_{l,l+1}$ on the entire spin chain.
Specifically, if we set the corresponding projector of the blockade as $P=\prod_{l=1}^{N}(\mathbb{I}-\Pi_l^{(j)}\otimes \Pi_{l+1}^{(-j)})$, then the blockade effect can be expressed as:
\begin{equation}\label{blocakdeandaverageofunitaryparts}
H(0)P =\frac{1}{2}[J^x(0)+J^x(\pi)]P=P J^x(0) P.
\end{equation}
Therefore none states located within the subspace defined by the projector $P$ can evolve out of this subspace. A detailed discussion about the mechanism of fragmentation and the proof of Eq.(\ref{blocakdeandaverageofunitaryparts}) can be found in Appendix A.

\begin{figure}
	\centering
	\includegraphics[width=\linewidth]{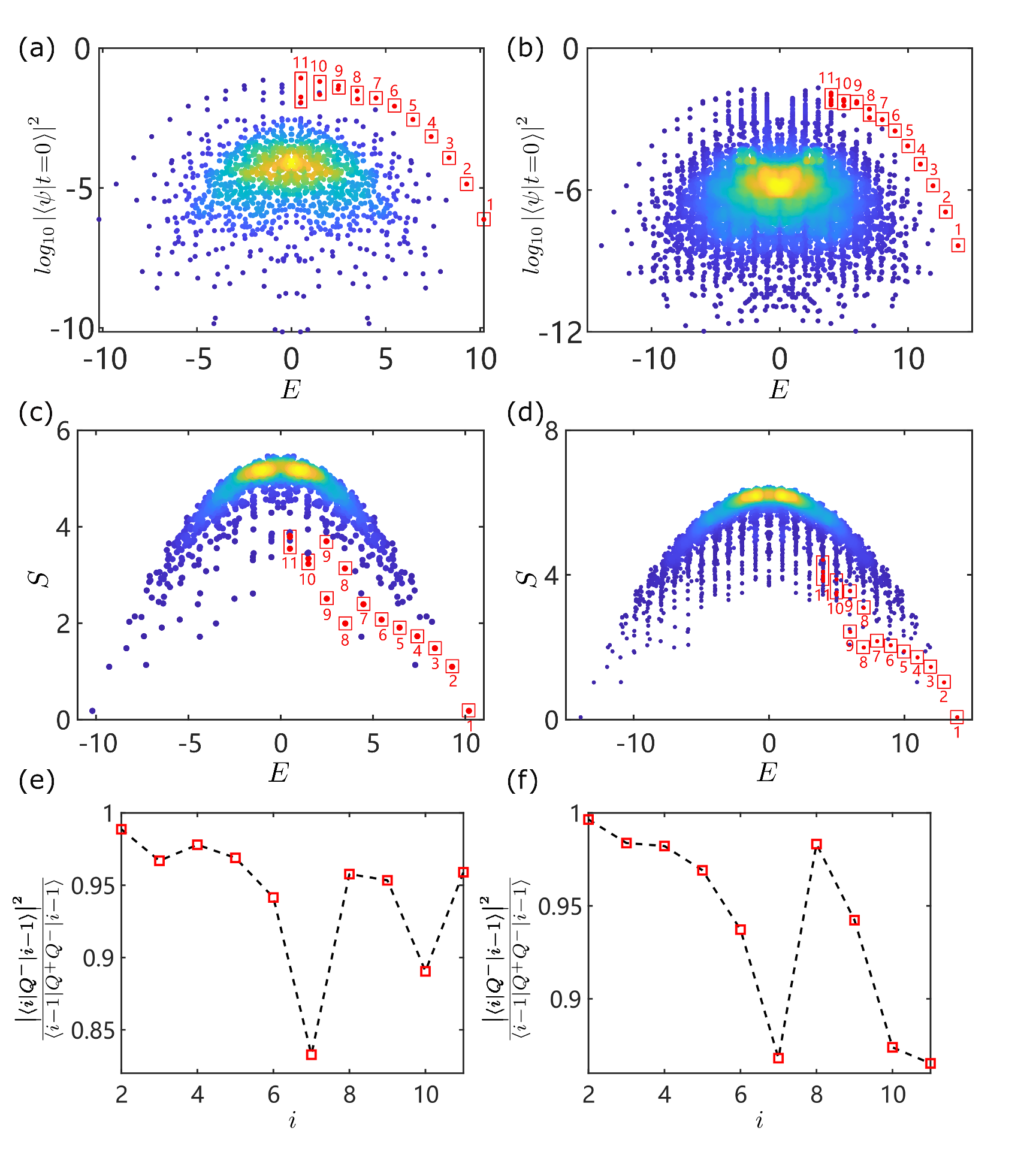}
	\caption{(a)-(d) The overlaps between eigenstates of the scar model with the initial states $\left|\psi(0)\right\rangle=\left|j,j,...,j\right\rangle_z$ and the relevant von Neumann entropies of these eigenstates.  Here (a) and (c) show the numerical results for a spin-$\frac{3}{2}$ chain, while the same calculations for a spin-$2$ chain are also plotted in (b) and (d), the Von Neumann entropies are calculated by taking half the chain as the subsystem $A$ throughout the entire work. The so-called scar towers become prominent with eigenstates possessing high overlaps with initial states and rather low entropies, compared with the buck of eigenstates governed by volume-law entropies. Here the calculation is done in spin chains with 7 spins, and we restrict the exact diagonalization to be done in the blockade subspace. It should be noted that in other eigen-subspace with $C\neq 0$, scar towers also appear with fewer numbers. In (a)-(d) we mark the highest 11 states of the scar states in red. In the middle part of the spectrum, when there are more than one eigenstates in a single rectangle, we appoint a specific superposition of the highly degenerate states as the scar state. In (e) and (f), corresponding to the spin-$\frac{3}{2}$ and spin-$2$ cases separately, we demonstrate the efficiency of $Q^-$ from Eq.(\ref{defineQca}) by a calculation of $\left|\left\langle i+1\right|Q^- \left| i \right\rangle \right|^2/\left\langle i \right|Q^+ Q^- \left| i \right\rangle$ with the horizontal axis being $i$, which is the serial number of scar states marked aside the red rectangles in sub-figure (a)-(d).}
	\label{fig: overlapandentropy}
\end{figure}

It is widely believed that the presence of QMBS is caused by the series of special scar eigenstates with approximately equal energy-spacing, or referred to as equally spaced eigenstates tower(QMBS tower)~\cite{scartower}. The entanglement entropy of these scar states deviates from that of the bulk eigenstates, and signifies the violation against the strong ETH. To confirm the existence of QMBS in our case with the $a=0$ Hamiltonian, in Fig.~\ref{fig: overlapandentropy}, we calculate the von Neumann entropy of eigenstates by deviding the chain into two equal parts for different spin sizes $j=3/2$ and $j=2$ in the $C=0$ subspace, the overlaps between these eigenstates and the initial state $|\psi(0)\rangle$ are also shown.
We find that such quasi-equally spaced scar eigenstates do exist, which can be separated from the bulk eigenstates by unusual low entropies, and high overlap with the initial states, which ensures the revival dynamics depicted in Fig.~\ref{fig: periodicalrevival}.

Although the current model shares a similar many-body scar phenomenon with the usual PXP model, the blockage interaction is quite different, leading to a completely different eigen-structure within Hilbert space. Specifically, for the $j=1/2$ system, the blockade term in PXP model prohibits adjacent atoms from being in the excited Rydberg state simultaneously. This means that patterns $\left|\uparrow,\uparrow \right\rangle_{l,l+1}$ are not allowed during the dynamical evolution.
However, in our case, if all adjacent spins are prohibited from occupying the states $\left|\uparrow,\downarrow \right\rangle_{l,l+1}$, namely $C=0$, then the only possible bases in this subspace are $\left|\downarrow,\cdots,\downarrow \right\rangle,\ \left|\uparrow,\cdots,\uparrow \right\rangle$, since PBC is employed. Therefore, the spin chain shall be dynamically static if we start with the chosen state $|\psi(0)\rangle$. Here $\left| \uparrow \right\rangle_{l}$ and $\left| \downarrow \right\rangle_{l}$ are the local spin-up and spin-down states in $z$-axis. Similarly in the $a\neq 0$ case, the system still can not be characterized as supporting QMBS though the perfect Loschmidt echo does appear, since the model shows integrability through the calculation of $\left\langle r \right\rangle$. For a detailed discussion, we suggest that readers refer to Appendix C.
Due to these reasons, in our case, the scar phenomena appears only when the spin size satisfies $j\geq1$.


Next, to gain a better understanding of this model, we prove that the spin-$1$ case of this model can be precisely mapped to a standard PXP model by defining new spin basis.
Specifically, when $a=0$, the Hamiltonian can be rewritten as
\begin{equation}\label{Hofrrll}
\begin{aligned}
      H(0)=\frac{1}{\sqrt{2}}\sum_{l=1}^{N}R_l\otimes R_{l+1}^2+  L_l^2 \otimes L_{l+1},
\end{aligned}
\end{equation}
where the matrix form of $L_l$ and $R_l$ under the local spin basis of $S_l^z$ $\{|m=1\rangle ,|m=0\rangle ,|m=-1\rangle \}$ are list as follows:
\begin{align}\label{lll}
  L_l=\left(\begin{array}{ccc}
      0 & 0 & 0 \\
      0 & 0 & 1 \\
      0 & 1 & 0
    \end{array}\right)_l, & \hspace{.5cm} R_l=\left(\begin{array}{ccc}
      0 & 1 & 0 \\
      1 & 0 & 0 \\
      0 & 0 & 0
    \end{array}\right)_l.
\end{align}
To map the above model to the usual PXP model, we redefine all these local basis $|m\rangle _l$ using composite spin-$1/2$ systems as
\begin{equation}\label{transformation}
\begin{aligned}
  & \left| m=1 \right\rangle_l=\left| \downarrow, \uparrow \right\rangle_{2l-1,2l}, \\
  & \left| m=0 \right\rangle_l=\left| \downarrow, \downarrow \right\rangle_{2l-1,2l},\\
  & \left| m=-1 \right\rangle_l=\left| \uparrow, \downarrow \right\rangle_{2l-1,2l}.
\end{aligned}
\end{equation}
The local operators $L_l$ and $R_l$ defined above can be rewritten as
\begin{equation}\label{eqRL}
    \begin{aligned}
        L_l & =\left( \left| \downarrow \right\rangle \left\langle \uparrow \right|_{2l-1}+\left| \uparrow \right\rangle \left\langle \downarrow \right|_{2l-1} \right)\otimes\left| \downarrow \right\rangle \left\langle \downarrow \right|_{2l}  \\
         & =\sigma_{2l-1}^x P_{2l}, \\
        R_l & = \left| \downarrow \right\rangle \left\langle \downarrow \right|_{2l-1} \otimes\left( \left| \downarrow \right\rangle \left\langle \uparrow \right|_{2l}+\left| \uparrow \right\rangle \left\langle \downarrow \right|_{2l} \right)  \\
         & =P_{2l-1} \sigma_{2l}^x, 
    \end{aligned}
\end{equation}
with $\sigma_l^{x,y,z}$ the Pauli operators and $P_l=\left| \downarrow \right\rangle \left\langle \downarrow \right|_l$ the projectors onto the spin-down states at lattice site $l$. Substituting the above results into Eq.(\ref{Hofrrll}), we finally have
\begin{equation}\label{HwHwithRatio}
\begin{aligned}
    H(0) & = \frac{1}{\sqrt{2}}\sum_{l=1}^{N}P_{2l-2} \sigma_{2l-1}^x P_{2l}+P_{2l-1} \sigma_{2l}^x P_{2l+1} \\
         & = \frac{1}{\sqrt{2}}H_{PXP}.
\end{aligned}
\end{equation}
This is exactly the PXP model in which quantum many-body scar was first experimentally discovered in Rydberg atomic chain~\cite{cjturner1,cjturner2}.
In this case, the chain length becomes twice that of the original spin-$1$ model, and the original blockade becomes a projection onto the subspace without adjacent Rydberg excited states.
It should also be noticed that the blockade of terms, i.e., $\left|\uparrow,\uparrow\right\rangle_{2l,2l+1}$, originates from the interaction term of the spin-$1$ model. Meanwhile, the other parts of the blockade terms $\left|\uparrow,\uparrow\right\rangle_{2l-1,2l}$ in Eq.(\ref{HwHwithRatio}) are completely prohibited due to the adoption of the specialized transformations shown in Eq.(\ref{transformation}), since such terms as $\left|\uparrow,\uparrow\right\rangle_{2l-1,2l}$ are not included in the transformations. Due to this reason, the spin model here can only be mapped to the PXP model in its blockade subspace.
Finally, the special initial state causing scar phenomenon in the spin chain model is $\left|\psi(0)\right\rangle=\left|1\rangle_1\otimes \cdots \otimes |1\right\rangle_N$, while in the PXP model, it represents the N\'eel state $\bigotimes_{l=1}^N | \downarrow, \uparrow \rangle_{2l-1,2l}$.
We notice that the transformation done here can also be found in an earlier work by Keita Omiya~\cite{projection}, where the research was initiated by considering the PXP Hamiltonian in the sight of Gutzwiller projection.


We emphasize that our approach is universal and can be generalized to the high-spin version of PXP model first discussed in ~\cite{highspinPXP}. To illustrate this, we consider a generalized definition of $\hat{C'}=\sum_{l=1}^{N}\Pi_l^{(a)}\otimes \Pi_{l+1}^{(b)}$ and its possibility for generating QMBS, here $\Pi_l^{(a)}=\left| a \right\rangle \left\langle a \right|_l, \Pi_l^{(b)}=\left| b \right\rangle \left\langle b \right|_l$, $\left| a\right\rangle_l$ and $\left| b\right\rangle_l$ denote arbitrary local states on the $l$-th site, the detailed discussion can be found in Appendix A. At first glance, it may seem that the work in ~\cite{highspinPXP} can not be straightly explained by our strategy, since we must let the blockaded nearest neighboring states $\left| a\right\rangle_l$ and $\left| b\right\rangle_l$ be orthogonal with each other (see in Appendix A). However, the high-spin version of PXP model in ~\cite{highspinPXP} does fit our description after we map two neighboring sites to a new logical lattice site. As an application of our strategy, we demonstrate this mapping briefly in Appendix A, which further demonstrates the universality of our method.

Generally speaking, the connection between blockade interactions and scar phenomenon is nontrivially unclear, although the possible relation has been investigated from both the perspective of Hilbert space fragmentation~\cite{scarforall} and the construction of blockade induced QMBS~\cite{systematic1d}.
For high-spin system,  we note that the constructing method used here differs from the strategy mentioned in ~\cite{QMBSsyzh,highspinPXP}, where blockade interactions take similar forms with the PXP model and are only introduced as a tool for generating QMBS.  Comparatively, our work provides another simple and systematic extension to construct scar model for large spin systems, which is quite universal and should benefit the investigation of such novel physics both in theory and experiment.

Next, we will demonstrate the intrinsic relation between $su(2)$ algebra and the spin chain based QMBS by proposing a $su(2)$-based ansatz for constructing the scars that we found numerically in subsequent sections.

\section{Proposing ansatz for constructing scar eigenstates with $a=0$.}

Various methods have been proposed to analytically construct scar states in order to gain more insights into the underlying physics.
Here we consider obtaining the scar eigenstates by constructing the Spectrum Generating Algebra(SGA)~\cite{creationoperators1,creationoperators2} of the system.
In this context, scar eigenstates are recognized as a series of states that can be obtained by successively acting on the seed state using so-called QMBS raising operators and lowering operators, which satisfy the special commutation relations with the Hamiltonian within a subspace of Hilbert space~\cite{SGA1,SGA2}.
For our case with $a=0$, the underlying raising and lowering operators of the spin models are written with the deformed '$su(2)$' operators:
\begin{equation}\label{defineQca}
Q^{\pm} (a=0)\equiv\frac{Q^y(a=0) \pm iQ^z(a=0)}{\sqrt{2}}.
\end{equation}
Here the operators $Q^{\alpha}(a)\equiv\frac{1+a}{2}J^{\alpha}(0)+ \frac{1-a}{2}J^{\alpha}(\pi)$, $\alpha=x,y,z$ are averaged collective spin operators just like the Hamiltonian itself defined in Eq.(\ref{eqhamiltonian}). These trial operators satisfy the following commutation relation similar to the standard $su(2)$ case which captures the intrinsic dynamical structure of the model:
\begin{equation}\label{sgacommutingrelation}
  \begin{aligned}
  & [H, Q^{\pm}]=\pm Q^{\pm}+i\frac{1}{\sqrt{2}}j\hat{R}, \\
  & \hat{R}=\sum_{l=1}^{N} \Pi_l^{(j-1)}\otimes \Pi_{l+1}^{(-j)}-\Pi_l^{(j)}\otimes \Pi_{l+1}^{(-j+1)}.
  \end{aligned}
\end{equation}
The proof of Eq.(\ref{sgacommutingrelation}) is given in Appendix B. This allows us to define the subspace $\mathcal{K}$ such that $\hat{R}\mathcal{K}=0$ is satisfied, and we have $\left[ H,Q^{\pm} \right]\mathcal{K}=\pm Q^{\pm}\mathcal{K}$.
The scar generating operators defined here commute with the counting operator $\hat{C}$, indicating that the relation holds in all the fragmented subspaces separately.
It should also be noticed that the relation displayed in Eq.(\ref{sgacommutingrelation}) deviates from those in standard SGA methods since the scar states are not always confined in the subspace $\mathcal{K}$. Therefore, this relation is not precise for the generating of scar states.
In Fig.~\ref{fig: overlapandentropy}(e) and ~\ref{fig: overlapandentropy}(f), we show numerically the performance of $Q^{\pm}$ employed here to generate scar states even though the exact SGA cannot be found analytically.
For instance, $Q^-$ acts pretty accurately as the lowering operator between every two nearest scar states, with at least $83$-percent fidelity mapping the higher energy states to lower ones.

With the ladder operator, we can approximately write down all the scar states given any one of them. For example, the ground state of the system can be approximated by projecting the ground state of the non-interacting collective spin operator $J^x(0)$ onto the $C=0$ space as
\begin{eqnarray}
    | GS \rangle_{app}= P\bigotimes_{l=1}^N| -j\rangle_l^x,
\end{eqnarray}
 where $\left|-j\right\rangle_l^x$ is the lowest eigenstate of the operator $S_l^x$. This approximation has an accuracy of over 99 percent in the case of our calculation with $j=1,3/2,2$ and $N=7$. Then the set of scar states can be approximated by
 \begin{eqnarray}
     \{ | GS \rangle_{app}, Q^+| GS \rangle_{app},\cdots,{Q^+}^{2Nj}| GS \rangle_{app}\}.
 \end{eqnarray}

For $j=1$, we transfer the aforementioned algebra into the PXP representation, and examine the validity of our strategy in comparison with the analysis done upon PXP model in reference~\cite{pxpsga}.
In the latter case, the definitions of the raising and lowering operators for scar states have been provided in the literature of \cite{pxpsga}, which take the following forms
\begin{equation}\label{sgapxpref}
  S_{\pi}^{\pm} (\alpha)=\frac{ Y_{\pi} \pm i \alpha Z_{\pi}}{2\sqrt{2}},
\end{equation}
with $\alpha \simeq 1/2$. They satisfy the relation $[H_{PXP},S_{\pi}^{\pm}]=\pm S_{\pi}^{\pm}+\hat{O}$, the specific form of $\hat{O}$ is not concerned here, which can be found in \cite{pxpsga}. Here $Y_{\pi}$ and $Z_{\pi}$ are termed as magnon operators carrying momentum $\pi$, and read
\begin{equation}\label{lll}
Z_{\pi}=\sum_{i=1}^{2N} (-1)^i \sigma_i^z, \hspace{0.5cm}
Y_{\pi}=\sum_{i=1}^{2N} (-1)^i P_{i-1} \sigma_i^y P_{i+1}.
\end{equation}
On the other hand, by using the mapping defined in Eq.(\ref{transformation}), we can naturally get the above PXP magnon operators from the averaged spin operators as:
\begin{equation}\label{lll}
\begin{aligned}
  Q^y(0)=\frac{J^y+U_{\pi} J^y U_{\pi}}{2}   &\longrightarrow \frac{1}{\sqrt{2}}Y_{\pi}, \\
  Q^z(0)=\frac{J^z+U_{\pi} J^z U_{\pi}}{2}=J^z   &\longrightarrow \frac{1}{2}Z_{\pi}.
\end{aligned}
\end{equation}
Therefore, the proposed creation and annihilation operators $Q^{\pm}(0)$ can be written as
\begin{equation}\label{sgapxpspin1}
  Q^{\pm}(0)=\frac{Q^y(0)\pm iQ^z(0)}{\sqrt{2}} \rightarrow \frac{Y_{\pi} \pm i\frac{1}{\sqrt{2}} Z_{\pi}}{2},
\end{equation}
which coincide with the form of Eq.(\ref{sgapxpref}), meaning that the operators $Q^{\pm}(0)$ and $S_{\pi}^{\pm}$ defined in these two different cases take very similar forms with slightly different coefficients ($\alpha \simeq 1/2$ and $1/\sqrt{2}$ respectively).
Therefore, the magnon operators emerged in PXP model can be understood through our definition of raising and lowering operators without the necessity of introducing $\pi$-magnon description.

\begin{figure}
  \centering
  \includegraphics[width=\linewidth]{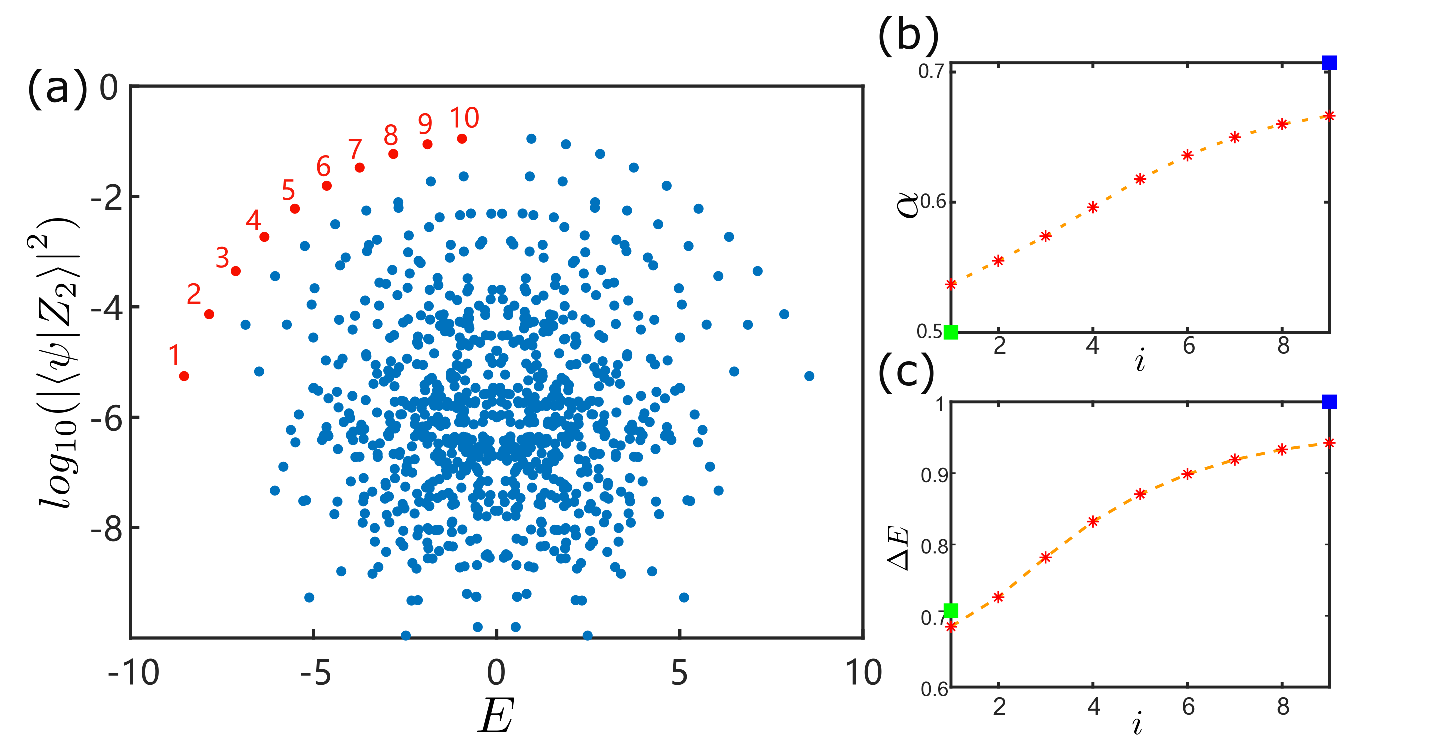}
  \caption{(a) Overlaps of eigenstates with the initial state $\left|\psi(0)\right\rangle$ calculated for the scar model in a 10-site spin-$1$ chain. Here we mark half of the scar states in red. It should be noted that the zero-energy scar state, which is in the middle of the spectrum, is missed since the calculation done here leaves out zero-energy eigenstates. (b) Minimized $\alpha_i$ from $f_i(\alpha)$, here the horizontal axis $i$ represents the serial number of scar states, as marked out in (a). (c) The energy spacings concerning serial numbers of the scar states. In both (b) and (c), theoretically anticipated values of $\alpha$ and the energy spacing $\Delta E$ are shown for both strategies discussed in the context. Here blue squares in the upper right corner represent the result of the theory we put forward in spin-$1$ model, while green squares in the left lower corner are the results from PXP models in previous studies~\cite{pxpsga}.}\label{fig: test of sga}
\end{figure}

The difference in the coefficient $\alpha$ can be explained by calculating the minimization of the relevant cost function defined below on every two adjacent scar states
\begin{equation}\label{lll}
\begin{aligned}
f_i(\alpha)=1-\frac{1}{2} &\left(  \frac{\left|\left\langle i+1\left|S_{\pi}^+(\alpha)\right|i\right\rangle\right|^2}{\left|\left\langle i \left|S_{\pi}^{-} (\alpha)S_{\pi}^+(\alpha)\right| i \right\rangle\right|} + \right. \\
 &\left. \hspace{.5cm} \frac{\left|\left\langle i\left|S_{\pi}^{-}(\alpha)\right| i+1\right\rangle\right|^2}{\left|\left\langle i+1\left|S_{\pi}^{+}(\alpha) S_{\pi}^{-}(\alpha)\right| i+1\right\rangle\right|} \right),
\end{aligned}
\end{equation}
here $\left|i\right\rangle$ denotes the $i$-th scar state counting from the left side of the spectrum. The optimal $\alpha_i$ is found such that $S^{\pm}(\alpha_i)$ can be viewed as the ideal creation and annihilation operators for the two neighboring scar states $|i\rangle$ and $|i+1\rangle$, namely $S^{+}(\alpha_i)|i\rangle \sim |i+1\rangle$ and $S^{-}(\alpha_i)|i+1\rangle \sim |i\rangle$.
In ~\cite{pxpsga}, only the ground state and the first excited state(1st and 2nd scar states) are considered and the optimal value of $\alpha_0$ is approximately $\alpha_0 \simeq 1/2$.
Here, we pick the lowest $L$ scar states, namely, the eigenstates from the negative part of the spectrum, as the probes to testify the SGA here and search for the optimum $\alpha$.
In Fig.~\ref{fig: test of sga}, we plot the numerically optimized $\alpha_i$ for neighboring pairs of scar states.
One can check that $\alpha_i$ takes different values for different pairs of adjacent scar states and tends to the limit case $1/\sqrt{2}\sim 0.707$ for growing $i$.
Therefore, the proposed creation and annihilation operators $Q^{\pm}(0)$ become more accurate in the middle of the entire spectrum.


It is also worth noticing that the level spacing predicted by these two different algebras above is also different by a factor of $\sqrt{2}$, which is the ratio between the Hamiltonian of the PXP model and the spin-$1$ model, as shown in Eq.(\ref{HwHwithRatio}).
In Fig.~\ref{fig: test of sga}, we also numerically calculate the energy spacings of different adjacent scar states.
One can see that the raising and lowering operators from ~\cite{pxpsga} work better in the edge of the spectrum, where the energy difference predicted fits the numerical results well, as indicated by the green squares in Fig.~\ref{fig: test of sga}.
However, in the middle of the spectrum, our strategy defined in Eq.(\ref{sgapxpspin1}) for spin-$1$ model can provide a more precise prediction of the energy spacing, which captures the behavior of the scar states better and is marked as blue squares in Fig.~\ref{fig: test of sga}.

\section{Universal Scenario With $a\neq 0$.}
The scar phenomenon discussed above can be extended to the general case with $a\neq0$ using similar methods.
Specifically, in the extreme case with $a=1$, the system becomes completely integrable and exactly solvable, which leads to non-thermalize dynamics of the system.
Therefore, it is expected that as $a$ decreases from $1$ to $0$, the system transitions smoothly from being completely integrable to completely non-integrable.
This allows us to discuss the emergence of quantum scar states under this transition.
We find that the entanglement entropies of eigenstates gradually become obeying volume-law, while the entropies of scar states remain lower than that. 

\begin{figure}
	\centering
	\includegraphics[width=\linewidth]{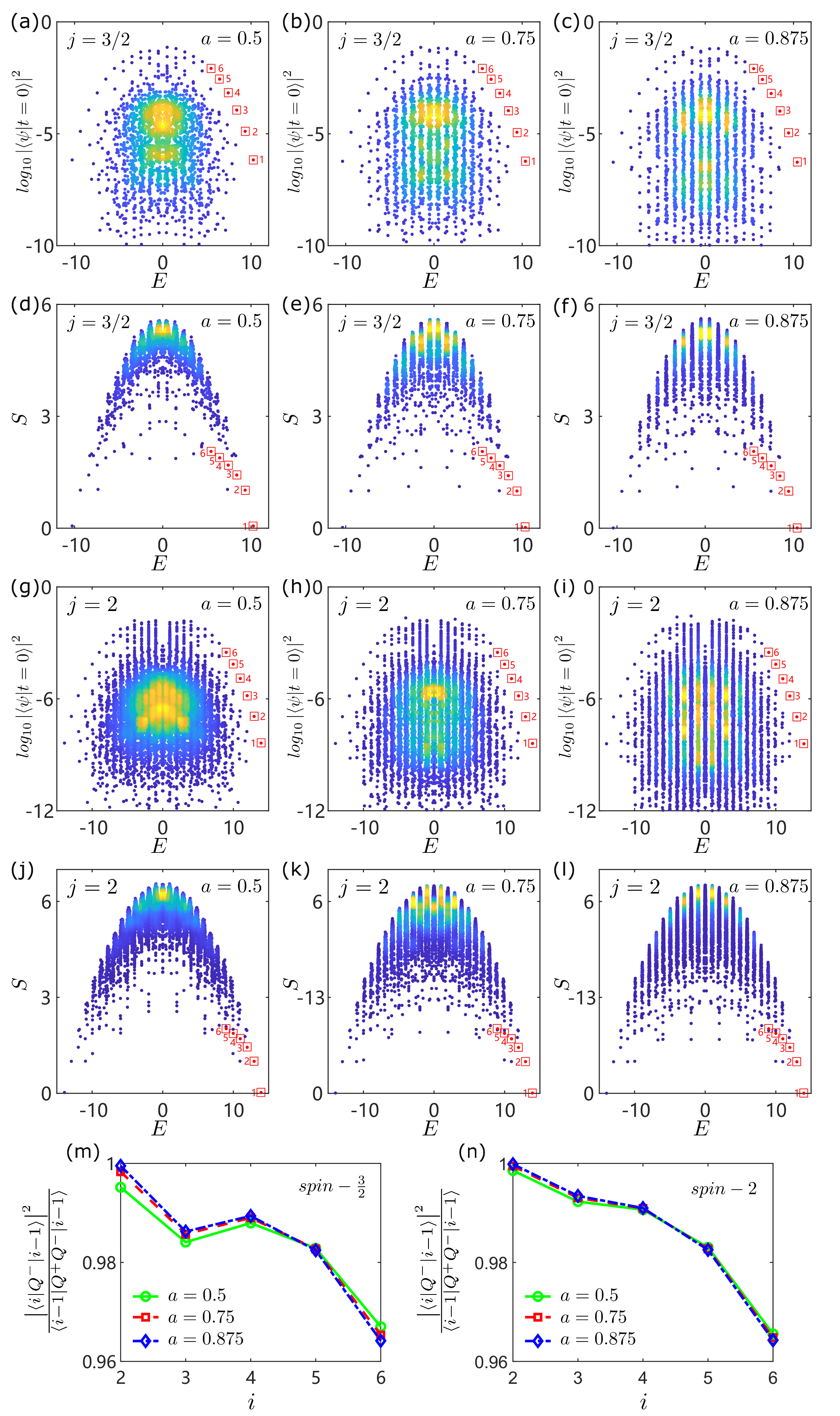}
	\caption{(a)-(l) The overlaps between the eigenstates of the scar model with the initial states and the relevant von Neumann entropies of these eigenstates for different parameters of $a$ and spin size $j$. The calculation is carried out through exact diagonalization of the translation-invariant lattice in the entire Hilbert space with the lattice size $N=7$. In (a)-(c), the overlaps of eigenstates with the initial state are displayed for spin-$\frac{3}{2}$ chains with $a=1-(\frac{1}{2})^k: k=1,\ 2,\ and\ 3$, while in (d)-(f) we calculate the corresponding bipartite entropies of these spin models, taking half the chain as subsystem $A$. In (g)-(l), the same calculations are carried out in the spin-$2$ model taking the same parameter $a$. The 'scar-like' towers remain obvious throughout the entire calculations. However, the bipartite entropies of eigenstates become lower in general, which is consistent with the fact that the system under consideration is no longer strictly non-integrable, as shown in Fig.~\ref{fig: calsforrmt}. The highest 6 'scar' states are marked in red squares, which are employed to testify the efficiency of SGA operators introduced in Eq.(\ref{sgaAneq0}) for spin-$\frac{3}{2}$ and spin-$2$ chains separately, as shown in (m) and (n).}
	\label{fig: aNeq0overlapandentropy}
\end{figure}

In Fig.~\ref{fig: periodicalrevival}, the periodical revival of the system under time-evolution with the appointed initial states, namely $\left|\psi(0) \right\rangle=\bigotimes_{l=0}^N |j\rangle_l^z$, has been shown for different $a$.
This indicates the robustness of scar states during the whole parameter settings.
In Fig.~\ref{fig: aNeq0overlapandentropy}, we also plot the overlaps of eigenstates with the initial states $\left|\psi(0)\right\rangle$ and the bipartite entropies of these eigenstates for different $a\neq 0$ with spin sizes $j=3/2$ and $j=2$ respectively, just as those depicted in Fig.~\ref{fig: overlapandentropy} with $a=0$ scenario.
The scar-like eigenstates can be spotted immediately due to their higher overlaps with initial states and relatively lower entropies.
Those 'scars' share the same intrinsic eigenstructure of scar eigenstates possessed by the $a=0$ case.
However, since the Hamiltonian gradually becomes more integrable-like and non-ergodic with the increasing parameter $a$ approximating $a=1$, as can be seen from Fig.~\ref{fig: calsforrmt}, they are not technically QMBS models.
Meanwhile, the von Neumann entropies of eigenstates also become lower in general with tower-like structures as $a$ increases, which shows an overall gradual deviation from the volume law in comparison with the $a=0$ case.

%

Thanks to the similar structure of the models discussed here, the corresponding spectrum-generating raising and lowering operators of scar states for general $a\neq 0$ can also be constructed as
\begin{equation}\label{sgaAneq0}
 Q^{\pm} (a)\equiv\frac{Q^y(a) \pm iQ^z(a)}{\sqrt{2}},
\end{equation}
with the operators $Q^{y,z}(a)\equiv\frac{1+a}{2}J^{y,z}(0)+ \frac{1-a}{2}J^{y,z}(\pi)$ defined in section V.
Therefore $Q^{\pm} (a)$ reduce to $Q^{\pm} (0)$ when $a=0$, and become ideal raising and lowering operators $J_x^{\pm}$ of the collective spin along the $x$-axis as $a \rightarrow 1$. The relevant commutation relation can also be verified as
\begin{equation}\label{wholeLiebracket}
[H(a), Q^{\pm}(a)]=\pm Q^{\pm}(a)+i\frac{1}{\sqrt{2}}(1-a^2)j\hat{R},
\end{equation}
with $\hat{R}$ the matrix defined in Eq.(\ref{sgacommutingrelation}). The deduction of Eq.(\ref{wholeLiebracket}) can be found in Appendix B. It is worth noticing that the commutation relation becomes exact when $a=1$, that is, at the edge of the circle on the parametric plane shown in Fig.~\ref{fig: Demonstration of the parametrical distribution of hamiltonian}, $Q^{\pm}(a=1)$ is exactly the generator of corresponding $su(2)$ algebra, which further guarantee the efficiency of the eigenstructure we constructed here.

In Fig.~\ref{fig: aNeq0overlapandentropy}(m) and \ref{fig: aNeq0overlapandentropy}(n), we also numerically verify the accuracy of the defined operators $Q^{\pm}(a)$ for neighbor scar states with different $a$ and spin size $j$. For each pair of neighboring scar states, the operator is capable of rather precisely performing the transition between the two states.
Here the quantity 
\begin{equation}\label{lll}
  \frac{\left|\left\langle i+1\left|Q^+(a)\right|i\right\rangle\right|^2}{\left|\left\langle i \left|Q^{-} (a)Q^+(a)\right| i \right\rangle\right|}
\end{equation}
captures the efficiency of the raising operator $Q^+(a)$, with $\left| i \right\rangle$ being the $i$-th scar states counting from right side of the spectrum, the calculation shows a slight difference of accuracy for varying $a$'s when mapping the highest 'scars' to the second highest ones, and as $a$ approaches $1$, the deviation becomes smaller and almost negligible, which is also consistent with the commutation relation defined in Eq.(\ref{wholeLiebracket}).

We note that the above discussion provides a new perspective for understanding the blockade-based scar states interpreted in the $a=0$ case. Basing on an altered $su(2)$ algebra, they are inherited from integrable models, i.e., the $a=1$ exactly solvable scenario, with quasi-equally spaced scars' eigenstructure.
This can be seen from the calculation of eigenstates' bipartite entropies and the inherent relationship between the scars' algebraic structure and the 'scar-like' ones, which is established by the formally unified commutation relation in Eq.(\ref{wholeLiebracket}).
Although the blockade interaction plays an important role in the formation of QMBS, our construction strategy  still ensures that the scar state has strong robustness, as changing parameters $a$ and $\theta$ can not sabotage the periodical behavior.

\section{Conclusions}
In summary, we have constructed the QMBS model by equally weighting two simple Hamiltonians satifying $su(2)$ algebra. By combining these two parts we find that there is another symmetry characterized by the operator $\hat{C}$ which leads to the emergence of weak fragmentation and blockade interactions. At the same time, the scar states originate from the eigen-structure hidden in the $su(2)$ algebra(a 'quasi-$su(2)$' algebra). By tunning the weight ratio, the model can transition from non-integrable QMBS model to exactly solvable non-chaotic one, where the scar states possess robustness against disturbance in the Hamiltonian parameters.
We have also analyzed the properties and explicit forms of these scar states both numerically and theoretically for different spin sizes $j$, demonstrating that the $\pi$-magnon type excitation from the PXP model can be viewed as a natural result of our assumption.
We note that the construction method used here is universal for introducing non-local symmetries like $\hat{C}$ in non-interacting systems, and can be generalized to construct a large class of 1D models that host fragmentation and QMBS, covering previous QMBS models in earlier works, such as~\cite{cjturner1,highspinPXP}.
On the other hand, as a simple toy model, this system exhibits various adjustable parameters such as parameter $a$ and spin size $j$. This also allows for the exploration of transitions from quantum integrable models to non-integrable ones, as well as from classical chaos to quantum chaos.

\begin{acknowledgments}
We would like to thank Shun-Yao Zhang for helpful discussions, and Xi-Wang Luo, Pan Gao for their assistence with the calculation. This work was funded by National Natural Science Foundation of China (Grants No. 11974334 and No. 11774332), and Innovation Program for Quantum Science and Technology (Grant No. 2021ZD0301200). XFZ also acknowledge support from CAS Project for Young Scientists in Basic Research (Grant No. YSBR-049).
\end{acknowledgments}

\bibliography{ref_note2}

\appendix

\section{Correspondence between the operator $\hat{C}$ and the blockade effect}
Hamiltonian $H(0)$ in the $C=0$ subspace can be viewed as the outcome of introducing blockade interaction into the non-interacting Hamiltonian. We will demonstrate this by giving a generalized proof of Eq.(\ref{blocakdeandaverageofunitaryparts}). 

To begin with, we generalize the counting operator $\hat{C}$ of the patterns $\left| j,-j\right\rangle_{l,l+1}$ and the unitary operator as:
\begin{equation}\label{newcountingop}
\begin{aligned}
  & \hat{C'}\equiv\sum_{l=1}^{N}p_l =\sum_{l=1}^{N}\Pi_l^{(a)}\otimes \Pi_{l+1}^{(b)}, \\
  & U'(\theta)\equiv e^{i\theta \hat{C'}}=\prod\limits_{l=1}^{N}\left[ \mathbb{I}-(1-e^{i\theta}) \Pi_l^{(a)}\otimes \Pi_{l+1}^{(b)} \right],
\end{aligned}
\end{equation}
where $\Pi_l^{(a)}=\left| a \right\rangle \left\langle a \right|_l$ and $\Pi_l^{(b)}=\left| b \right\rangle \left\langle b \right|_l$ are local projectors, $\left| a \right\rangle_l$ and $\left| b \right\rangle_l$ are arbitrary local states of the $l$-th site, $p_l=\Pi_l^{(a)}\otimes \Pi_{l+1}^{(b)}$ is a quasi-local projector. Therefore, $\hat{C'}$ is a counting operator of the patterns $\left| a,b \right\rangle_{l,l+1}$ in the whole chain with periodic boundary condition. Then we demonstrate that any Hamiltonian of the form Eq.(\ref{a0Hamiltonian}) commutes with the corresponding $\hat{C'}$ given $_l\left\langle a|b\right\rangle_l=0$.  This can be verified by expressing $H'(0)$ as:
\begin{equation}\label{lll}
\begin{aligned}
  H'(0) & =\frac{J^x(0) +J^x(\pi)}{2} \\
        & =\frac{1}{2}\sum_{l=1}^{N}\left( S_l^x + e^{i\pi(p_{l-1}+p_{l})}S_l^x e^{-i\pi(p_{l-1}+p_{l})}\right).
\end{aligned}
\end{equation}
The commutation of $H'(0)$ and $\hat{C'}$ writes:
\begin{equation}\label{commuex}
\begin{aligned}
  & [H'(0), \hat{C'}]= \\
  & \frac{1}{2}\sum_{l=1}^{N}[ \underbrace{S_l^x}_{I} + \underbrace{e^{i\pi(p_{l-1}+p_{l})}S_l^x e^{-i\pi(p_{l-1}+p_{l})}}_{II}, p_{l-1}+p_{l} ].
\end{aligned}
\end{equation}
For part II of the commutation relation we have:
\begin{eqnarray}
    && [e^{i\pi(p_{l-1}+p_{l})}S_l^x e^{-i\pi(p_{l-1}+p_{l})}, p_{l-1}+p_{l} ] \nonumber\\
  &&=  \left( 1-2(p_{l-1}+p_{l}) \right) \left[S_l^x, p_{l-1}+p_{l} \right] \left( 1-2(p_{l-1}+p_{l}) \right) \nonumber\\
  &&=  -\left[S_l^x , p_{l-1}+p_{l} \right], \label{lll}
  \end{eqnarray}
which cancels with part I, therefore we have $[H'(0),\hat{C'}]=0$. 

Next, we show that in the $C'=0$ subspace, $H'(0)$ satisfies Eq.(\ref{blocakdeandaverageofunitaryparts}). By defining $\left| a' \right\rangle_l =S_l^x \left| a \right\rangle_l, \left| b' \right\rangle_l =S_l^x \left| b \right\rangle_l$, and $\hat{X}^{(a,a')}_{l}=\left|a \right\rangle \left\langle a' \right|_{l}+\left|a' \right\rangle \left\langle a \right|_{l}$, the Hamiltonian can be rewritten as:
\begin{equation}\label{h0detailedab}
  \begin{aligned}
  H'(0) & =\sum_{l=1}^{N}S_l^x -\hat{X}^{(a,a')}_{l} \otimes \Pi_{l+1}^{(b)}-\Pi_l^{(a)}\otimes\hat{X}^{(b,b')}_{l+1} \\
  &+2( _l\left\langle a|a'\right\rangle_l+_l\left\langle b|b'\right\rangle_l )\Pi_l^{(a)}\otimes\Pi_{l+1}^{(b)} \\
       & +2 (_l\left\langle b|a'\right\rangle_l) \Pi_{l-1}^{(a)} \otimes \left| b \right\rangle\left\langle a \right|_{l} \otimes \Pi_{l+1}^{(b)}+h.c.. 
  \end{aligned}
\end{equation}
Since the last two terms vanish after acting the projector $P'=\prod_{l=1}^{N}(\mathbb{I}-\Pi_l^{(a)}\otimes\Pi_{l+1}^{(b)})$ on the right side of $H'(0)$, the remaining parts of $H'(0)$ match the blockade formation from Eq.(\ref{blocakdeandaverageofunitaryparts}):
\begin{eqnarray}
  &&\left(\sum_{l=1}^{N}S_l^x -\hat{X}^{(a,a')}_{l} \otimes \Pi_{l+1}^{(b)}-\Pi_l^{(a)}\otimes\hat{X}^{(b,b')}_{l+1}\right)P' \nonumber \\
  && \hspace{2.cm} = P'J^x(0)P'. \label{lll}
\end{eqnarray}
In the typical case with $\left| a \right\rangle_l =\left| j \right\rangle_l, \left| b \right\rangle_l =\left| -j \right\rangle_l$, and $j\geq 1$, the last two terms in Eq.(\ref{h0detailedab}) vanish for any values of $C$. Therefore, the pattern $\left| j,-j\right\rangle_{l,l+1}$ is frozen in all the subspaces labeled by $C$. Hence the Hilbert space further fragments into smaller subspaces with fixed $\left| j,-j\right\rangle_{l,l+1}$.

We note that the above deduction does not concern the specific form of $S_l^x$, in other words, it still holds for general non-interacting Hamiltonian $H_0=\sum_{l=1}^{N}h_l$. Hamiltonians constructed in this way possess the symmetry of $\hat{C'}$, and may cause fragmentation when $\left| a \right\rangle_l$ and $\left| b \right\rangle_l$(or the projectors $\Pi_l^{(a)}$ and $\Pi_l^{(b)}$) satisfy the following conditions:
\begin{equation}\label{condforF}
\begin{aligned}
  & _l\left\langle a|b \right\rangle_l=0,\  _l\left\langle a\right| h_l \left|b \right\rangle_l=0. \\
  & (\mbox{or }\ \Pi_l^{(a)}\Pi_l^{(b)}=0,\ \Pi_l^{(a)}h_l\Pi_l^{(b)}=0.)
\end{aligned}
\end{equation}
Generally speaking, fragmentation does not necessarily leads to QMBS. Nonetheless we will demonstrate that the higher-spin version of PXP model brought up in ~\cite{highspinPXP} also fits our description after mapping two neighboring sites into a single logical site. The Hamiltonian in ~\cite{highspinPXP} reads:
\begin{equation}\label{origHhpxp}
\begin{aligned}
      H_{hPXP} & =\mathcal{P} \sum_{l=1}^{N}S_l^x \mathcal{P}, \\
      \mathcal{P} & =\prod_{l=1}^{N}(\mathbb{I}_{l,l+1}-Q_l\otimes Q_{l+1}), \\
      Q_l & =\mathbb{I}_l-\Pi^{(0)}_l=\sum_{k=1}^{2s}\left| k \right\rangle_l\left\langle k \right|,
\end{aligned}
\end{equation}
where $\{\left| 0 \right\rangle_l,\left| 1 \right\rangle_l,\cdots,\left| 2s \right\rangle_l\}$ are the sets of $2s+1$ eigenstates of $S_l^z$ on the $l$-th site, and $\Pi^{(0)}_l=\left| 0\right\rangle \left\langle 0\right|_l$. Since $\mathcal{P}$ prohibits neighboring spins from both being in the space spanned by $Q_l$, the only allowed states in two neighboring sites are:
\begin{equation}\label{lll}
\begin{aligned}
  \{\left| 0,0 \right\rangle_{2l-1,2l}, & \left| 0,1 \right\rangle_{2l-1,2l},\cdots,\left| 0,2s \right\rangle_{2l-1,2l}, \\
   & \left| 1,0 \right\rangle_{2l-1,2l},\cdots,\left| 2s,0 \right\rangle_{2l-1,2l}\},
\end{aligned}
\end{equation}
we assign a new set of sites with half length and states defined as:
\begin{equation}\label{lll}
 \begin{aligned}
 & \left| a_k \right\rangle_{l}=\left| 0,k \right\rangle_{2l-1,2l}, \ k=1,2,\cdots,2s \\
 & \left| b_k \right\rangle_{l}=\left| k,0 \right\rangle_{2l-1,2l}, \ k=1,2,\cdots,2s \\
 & \left| c \right\rangle_{l}=\left| 0,0 \right\rangle_{2l-1,2l},
 \end{aligned}
\end{equation}
then we define new projector as:
\begin{equation}\label{lll}
 \begin{aligned}
 & \hspace{0.3in} \mathcal{P}' =\prod_{l=1}^{N/2}(\mathbb{I}_{l,l+1}-W_l\otimes V_{l+1}), \\
 & W_l=\sum_{k=1}^{2s}\left| a_k \right\rangle_{l}\left\langle a_k \right|, \quad\quad
   V_l=\sum_{k=1}^{2s}\left| b_k \right\rangle_{l}\left\langle b_k \right|,
 \end{aligned}
\end{equation}
and the non-interacting Hamiltonian as:
\begin{equation}\label{lll}
 \begin{aligned}
  H_0 & =\sum_{l=1}^{N/2}h_l \\
      & =\sum_{l=1}^{N/2}S_{2l-1}^{x}\otimes \left| 0\right\rangle_{2l}\left\langle 0\right|+\left| 0\right\rangle_{2l-1}\left\langle 0\right|\otimes S_{2l}^{x}.
  \end{aligned}
 \end{equation}
Then the Hamiltonian $H_{hPXP}$ is equivalent to $\mathcal{P}'H_0\mathcal{P}'$, and this Hamiltonian can be described by:
\begin{equation}\label{mappedH}
 \begin{aligned}
  & H'=\frac{H_0+U(\pi) H_0 U(\pi)}{2}, \\
  & U'(\theta)= e^{i\theta \hat{C}}=\prod_{l=1}^{N/2}(\mathbb{I}_{l,l+1}-2W_l\otimes V_{l+1}), \\
  & \hat{C}'=\sum_{l=1}^{N/2}W_l\otimes V_{l+1} .
 \end{aligned}
\end{equation}
Since it is easy to verify that projectors $W_l$ and $V_l$ satisfy the condition in Eq.(\ref{condforF}) as $W_lV_l=0$ and  $W_l h_l V_l=0$. Therefore we have:
\begin{equation}\label{lll}
  \begin{aligned}
   [ H',\hat{C}' ]&=0, \\
   H'\mathcal{P}'&=\mathcal{P}'H_0\mathcal{P}',
  \end{aligned}
\end{equation}
which means that Hamiltonian $H'$ possess symmetry $\hat{C}'$ and is equivalent to $H_{hPXP}$ in the $C'=0$ blockaded subspace.  Therefore, Hamiltonian $H'$ fits our strategy of defining a QMBS model and shares some key properties mentioned in the main text. For example, it also supports QMBS in the $\hat{C}'=0$ subspace and causes fragmentation.

\section{Deduction of the commutation relation between $H(\theta,a)$ and the raising and lowering operators}

The commutation relation in Eq.(\ref{wholeLiebracket}) is important for constructing the scar states and further understandings of the models' eigen-structure, here we present a proof of Eq.(\ref{wholeLiebracket}) for intrigued readers to follow:
\begin{equation}\label{lll}
  [H(a),Q^{\pm}(a)]=\pm Q^{\pm}(a)+i\frac{1}{\sqrt{2}}(1-a^2)j\hat{R},
\end{equation}
where $H(a)$ is the Hamiltonian defined in Eq.(\ref{eqhamiltonian}) and $Q^{\pm}(a)=[(1+a)J_x^{\pm}+(1-a)J_x^{\pm}(\pi)]/2$ are the ladder operators defined in Eq.(\ref{sgaAneq0}). The left side of the above commutation can be calculated as:
\begin{eqnarray}
    &&[H(a),Q^{\pm}(a)] \nonumber \\
  = && [\frac{1+a}{2}J^x+\frac{1-a}{2}U_{\pi}J^xU_{\pi},\frac{1+a}{2}J_x^{\pm}+\frac{1-a}{2}U_{\pi}J_x^{\pm}U_{\pi}] \nonumber\\
  = && \pm (\frac{(1+a)^2}{4}J_x^{\pm} + \frac{(1-a)^2}{4}U_{\pi}J_x^{\pm}U_{\pi}) \nonumber\\
    && \hspace{0.2in}+\frac{1-a^2}{4}([J^x,U_{\pi}J_x^{\pm}U_{\pi}]+U_{\pi}[J^x,U_{\pi}J_x^{\pm}U_{\pi}]U_{\pi}) \nonumber\\
  = && \pm (\frac{1+a}{2}J_x^{\pm} + \frac{1-a}{2}U_{\pi}J_x^{\pm}U_{\pi}) \nonumber\\
    && \hspace{0.2in}+ \frac{1-a^2}{4}\{U_{\pi},[J^x, U_{\pi}J_x^{\pm}U_{\pi}-J_x^{\pm}] \}U_{\pi} \nonumber\\
  = && \pm Q^{\pm}(a)  + \frac{1-a^2}{4}\{U_{\pi},[J^x, U_{\pi}J_x^{\pm}U_{\pi}-J_x^{\pm}] \}U_{\pi},  \label{expandcommu}
\end{eqnarray}
here the curly braces represent the anti-commutator $\{A,B\}=AB+BA$. Since the spin-$z$ operator $J^z$ remains unchanged under the unitary transformation $U_{\pi}$, we have:
\begin{eqnarray}
    &&U_{\pi}J_x^{\pm}U_{\pi}-J_x^{\pm} = \frac{1}{\sqrt{2}}(U_{\pi}J_y U_{\pi}-J_y) \nonumber\\
  && \quad\quad\quad =i\sqrt{j}\sum_{l=1}^{N} ( \left| j \right\rangle\left\langle j-1\right|_{l} \otimes \Pi_{l+1}^{(-j)} \nonumber\\
   && \quad\quad\quad \quad - \Pi_{l}^{(j)} \otimes \left| -j \right\rangle\left\langle -j+1\right|_{l+1})+h.c.. \label{Rcore}
\end{eqnarray}
Substituting Eq.(\ref{Rcore}) back into Eq.(\ref{expandcommu}), we have:
\begin{equation}\label{lll}
  \begin{aligned}
    & \frac{1-a^2}{4}\{U_{\pi},[J^x, U_{\pi}J_x^{\pm}U_{\pi}-J_x^{\pm}] \}U_{\pi} =i\frac{1}{\sqrt{2}}(1-a^2)j \hat{R}, \\
    & \hspace{0.2in}\hat{R}=\sum_{l=1}^{N} \Pi_l^{(j-1)}\otimes \Pi_{l+1}^{(-j)}-\Pi_l^{(j)}\otimes \Pi_{l+1}^{(-j+1)}. 
  \end{aligned}
\end{equation}
In the last step we have utilized the fact that $U_{\pi}$ flips the terms that contain $\left| j,-j \right\rangle_{l,l+1}$(or $_{l,l+1}\left\langle j,-j \right|$) and leaves the rest unchanged whenever it acts on the left (or right) side of a matrix.

\section{Discussion about the spin-$\frac{1}{2}$ case}

\begin{figure}
  \centering
  \includegraphics[width=7cm]{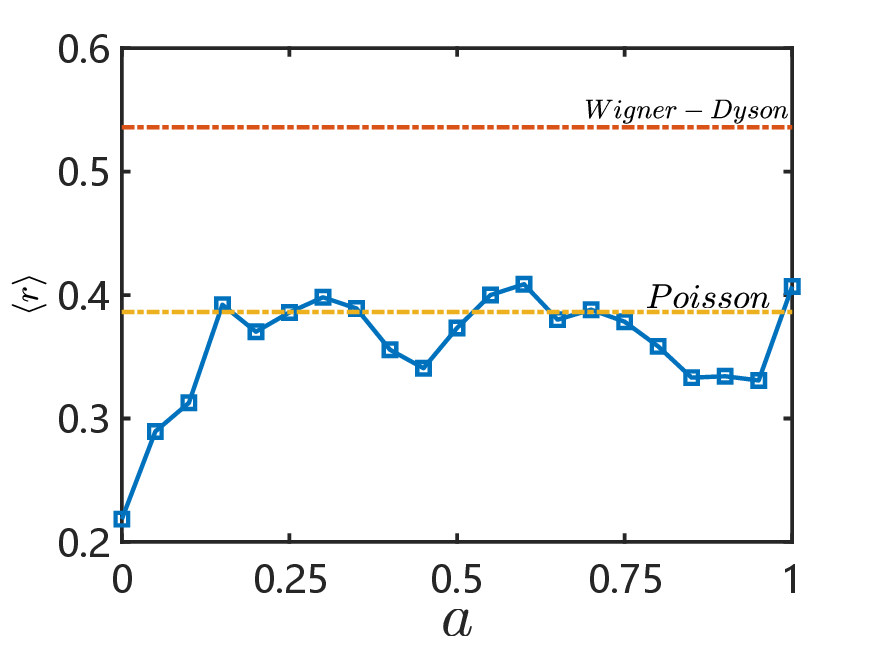}
  \caption{The average value of $r_i=\min(s_i,s_{i+1})/\max(s_i,s_{i+1})$ calculated for spin-$1/2$ system with parameter $a$ varying from $0$ to $1$. The calculation is carried out in a $12$ sites chain. The system displays integrability throughout the entire region of parameter $a$. }\label{fig: rmeanforspin05}
\end{figure}

Although the $spin-\frac{1}{2}$ model with $a\neq 0$ displays a perfect revival of overlaps with the initial state, this does not qualify it for a QMBS model, since the calculation of its energy level spacing implies integrability, as shown in Fig.~\ref{fig: rmeanforspin05}.

To understand the periodical revival in this case, we need to find the exact eigenstates that are responsible for such dynamical behavior. The Hamiltonian in Eq.(\ref{eqhamiltonian}) can be rewritten for spin-$1/2$ case
\begin{eqnarray}
    H(a) & =&\sum_{l=1}^{N}\frac{1}{2}\sigma_l^x+\frac{a-1}{2} \left(\sigma_l^x \otimes\left|\downarrow \right\rangle\left\langle \downarrow \right|_{l+1} + \left|\uparrow \right\rangle\left\langle \uparrow \right|_{l} \otimes \sigma_{l+1}^x \right)   \nonumber \\
   & =&\frac{a}{2} \sum_{l=1}^{N}\sigma_l^x+\frac{a-1}{4}\sum_{l=1}^{N}(\sigma_l^z \sigma_{l+1}^x-\sigma_l^x \sigma_{l+1}^z). \label{lll}
\end{eqnarray}
Although it is hard to analytically solve this Hamiltonian, we have been able to find a series of eigenstates causing the periodical revival
\begin{equation}\label{spinhalfstates}
  \{ \left| \phi \right\rangle, Q_{1/2}^+\left| \phi \right\rangle, (Q_{1/2}^+)^2\left| \phi \right\rangle,\cdots ,(Q_{1/2}^+)^N\left| \phi \right\rangle \},
\end{equation}
where $\left| \phi \right\rangle=\bigotimes^N_{l=1} |\downarrow\rangle_l^{x}$, and $\left|\downarrow\right\rangle^x_l$ is the $l$-th spin-down states in $x$-axis. $Q_{1/2}^{\pm}\equiv \sum_{l=1}^{N}S_l^{\pm}=\sum_{l=1}^{N}\frac{\sigma_l^y\pm i\sigma_l^z}{2\sqrt{2}}$ is the ladder operator for collective spins, which satisfies the following commutation relation:
\begin{equation}\label{lll}
  \begin{aligned}
  [H(a),Q_{1/2}^{\pm}] & =\pm a Q_{1/2}^{\pm} \\
  & \mp i\frac{a-1}{\sqrt{2}}\sum_{i=1}^{N}(S_i^+ S_{i+1}^- -S_i^- S_{i+1}^+),
  \end{aligned}
\end{equation}
with the last term vanishing when acting on the states from Eq.(\ref{spinhalfstates}). Therefore $Q_{1/2}^{\pm}$ are the exact ladder operators in the space $\mathcal{K'}$ spanned by the states in Eq.(\ref{spinhalfstates}). Since $\left| \phi \right\rangle$ is an eigenstate of $H(a)$, those states in Eq.(\ref{spinhalfstates}) are all eigenstates of $H(a)$ with energy interval $\Delta E=a$. Therefore, the periodical dynamics starting from $\left|\psi(0)\right\rangle=\bigotimes^N_{l=1} |\downarrow\rangle_l^{z}$  has a time period inversely proportional to $a$, since this initial state is completely in the subspace $\mathcal{K'}$. In the $a=0$ case, the states in $\mathcal{K'}$ are all degenerate, which means $\left|\psi(0)\right\rangle$ itself becomes an eigenstate of $H(a=0)$ and remains dynamically static. We can see both arguments confirmed in Fig.~\ref{fig: periodicalrevival}.

\end{document}